\documentstyle[ ]{l-aa}

\begin{document}

\thesaurus{08:03:4, 08:09:2, 08:11:1,}

\title{The Castor Moving Group}

\subtitle{The age of Fomalhaut and Vega\thanks{Based
on  observations collected by the Hipparcos satellite}}

\author{D.~Barrado y Navascu\'es \inst{1,2}          }

\offprints{D.~Barrado y Navascu\'es, dbarrado@cfa.harvard.edu}

\institute{MEC/Fulbright Fellow at the Smithsonian Astrophysical
Observatory, 
        60 Garden St, Cambridge, MA 02138, USA.
        \and
           Real Colegio Complutense at Harvard University,
        Trowbridge St, Cambridge, MA 02138, USA.}                     
  
\date{Received date, ; accepted date, }

\maketitle

\begin{abstract}

We have recomputed the kinematic properties of several of dozens nearby
stars, to try to verify if Castor ($\alpha$ Gem) indeed has a cohort
of stars sharing the same space motion and age. We used  kinematics,
the location of the stars in Color-Magnitude Diagrams, their 
lithium abundances, and
their activity, to establish that the moving group seems to be real and
to reject several stars which were believed to be part of this group.
 Of the initial 26 stars, we show that probably only
16 stars are physically associated.

The moving group contains several A spectral type stars. Among them, 
Vega and Fomalhaut, two of the prototypes of the $\beta$ Pic type stars.
If these stars are coeval, their different levels of  IR emission
 suggest that the time scale for the formation
of planets is not universal. Due to the age of the group, 
these IR excesses would appear as a consequence
of collisions and sublimation of larger bodies and they would not arise from 
protoplanetary structures.

Since this association includes several late spectral type stars, we used
their properties to estimate their age and,  therefore, the age of
the group and that of Vega and Fomalhaut. Our estimate for that age is
200$\pm$100 Myr.


\end{abstract}

\section{Introduction}

It was realized a few years ago (Anosova \& Orlov 1991)
 that the multiple system
ADS~6175, which contains  three spectroscopic binaries (Castor A,
Castor B  and YY Gem), could be just the most prominent members of 
a moving group. This moving group would include about 18 stars,
having spectral types between A1~V and M6~Ve, including
 one of the prototypes of the $\beta$ Pic stars.

The $\beta$ Pic type stars show IR excesses, which are
 associated the  the presence of
 circumstellar dusty disks.  The prototypes for this class of stars
 are $\beta$ Pic, Vega and Fomalhaut. 
  As a result of their proximity and brightness, these
three stars have been studied in great detail, and estimates
of the  mass, radial distribution and structure, dust-grain
properties of their disks have been made (e.g., Backman \& Paresce 1993). 
It is generally believed that these circumstellar dusty disks are 
either the direct descendents of T Tauri disks or the secondary
products  of the planet formation process.
 Knowledge of the ages of $\beta$ Pic stars is therefore
one of the keys to understand the formation and
evolution of their disks.   However, there is only a handful
of estimates of  the
ages of these systems, and no member of this class has yet been
detected with certainty in  an open cluster (although see Backman, Angelova
\& Stauffer 1998).  
All  three prototypes are A stars, and estimates from their
post-ZAMS evolution give approximate ages of 100, 200, and 400 Myr
 for $\beta$ Pic, Fomalhaut and Vega, respectively (Backman and
Paresce 1993). 
Recently, Barrado y Navascu\'es et al. (1997a) 
estimated an age of 200$\pm$100 My. for Fomalhaut, 
 based on a number of properties (X-ray emission, 
rotation, lithium abundance, isochrones) of its physical companion  GL879.
Another relevant star belonging to this class, HR4796A, has been studied
in a similar way by Stauffer et al. (1995). These authors derived an age of
 8$\pm$2 Myr.
Lately, Holland et al. (1998) have imaged the  thermal emission
from the disks of the three prototypes at  submillimeter wavelengths. They 
concluded that any Earth--like planet must have already  formed.
Moreover, the central holes found by Holland et al. (1998)
 in the orbit of Fomalhaut and by Jayawardhana et al. (1998) and Koerner et
 al. (1998) in the case 
of HR4796A, support the idea of the early formation of planets.

In this paper, we examine the proper motions of several nearby stars.
Based on Hipparcos data, we show that, indeed, Castor shares its 
Galactic motion with other stars, including Fomalhaut and Vega
and several late spectral type stars.
This fact allows us to derive the age of the group.

\section{The  selection of the sample}

A few years ago,
Barrado y Navascu\'es et al. (1997a) determined the age of one of the
prototypes of the $\beta$ Pic class, Fomalhaut, based on the 
properties of a physical late spectral type companion, GL879. They noted
 that Anosova \& Orlov (1991) had suggested that both stars could be part of a 
moving group which would include Castor A$\+$B and YY Gem.
However, uncertainties at that time were too large to derive any firm 
conclusion.
The wealth of data obtained by Hipparcos allows a more precise determination
of  distances, positions, and proper motions. Using these new values,
 we have recomputed the Galactic velocities for the Anosova \& Orlov (1991)
sample, to try to verify whether  there is a moving group or not.

In addition to the 18 possible members studied by Anosova \& Orlov (1991),
 we have included
several stars from two different sources. The first source provides  the
kinematics  of several types of cool dwarfs in the solar neighborhood
(Soderblom 1990). The second source is a large study which searched
 for kinematic groups in the solar neighborhood (Agekyan \& Orlov 1984).
These latter authors  computed the 
Galactic velocities for more than 1,000 stars selected from the Gliese Catalog,
and suggested  the existence of several moving groups. 
From these two sources, we have selected a total of 26 stars,
 and  we have re-analyzed their kinematic properties.
Table 1 lists our sample of stars. The first column contains the common
name for  well known stars, whereas the other three columns 
provide the numbers in the Gliese, HD and Hipparcos/Tycho catalogs, 
respectively.

A Color-Magnitude Diagram of the whole sample is shown in Figure 1, 
where the absolute visual magnitude is plotted versus the 
(B--V) color index. A Schmidt-Kaler (1982) Zero Age Main Sequence
(ZAMS) 
is also included as a solid line. The group includes  eight A stars with 
different levels of IR emission, as well as  five dF, two dG, three dK and
seven dM. Note that we have not performed
 a systematic search for possible members
(either using the whole Gliese Catalog or taking advantage of the Hipparcos
database) and, therefore, our sample is not complete and may be biased.

Table 2 contains additional information about the sample. We list the
apparent and absolute visual magnitudes, (B--V) and (V--I) --cousin--
colors and, when available, the equivalent width of
H$\alpha$ --EW(H$\alpha$)--, X-ray luminosity and
lithium abundance.

\section{Membership in the moving group}

\subsection{The kinematic criterion}

Based on the positions, parallaxes, and proper motions
 provided by the  Hipparcos database, and the radial velocities (SIMBAD
database and references therein), we have computed the kinematic properties
of our sample of stars. Table 3 shows the numbers in the Gliese Catalog, the 
positions, parallaxes, proper motions, radial velocities, and 
the Galactic velocities (U, V and W), with their associated errors for
all these quantities.
Galactic velocities and errors were computed following the description provided
by Johnson \& Soderblom (1987) and follow the right-handed coordinate
system (positive toward the Galactic center, Galactic rotation and North
Galactic Pole). Since the time span for the Hipparcos mission was short, the
proper motions based on Hipparcos data
might be not accurate. For this reason, we have also computed
the Galactic velocities using  the proper motion values from the PPM 
(Roeser \& Bastian 1988; Bastian \& Roeser 1993; Roeser \& Bastian 1994),
and  the Hipparcos positions and parallaxes.
 Except in very few cases (e.g. GL351, GL521.2, Castor itself),
 the discrepancies   do not affect the  conclusions. 
However, we note that the differences between the Hipparcos and the PPM
proper motions differ by more than the quoted uncertainties for all stars 
in the sample. Narayanan \& Gould (1998) have shown that Hipparcos 
proper motions are consistent with its parallaxes (at least in the direction
toward the Hyades), an indication that there are not systematics errors 
due to the short baseline. The derived V component of the Galactic velocities
 also differ by more than the nominal uncertainties in 30\% of the sample.
This fact indicates that some stars catalogued as members of the moving group
could be spurious members.

We have made a special effort to use the most accurate radial
velocities. Primarily, we selected the values from the WEB catalog
(Duflot et al. 1995), a compilation of three different works
(Wilson 1963; Evans 1978; Batten et al. 1989). We also provide other values
quoted in the literature. Only in three cases (GL426, GL564.1 and GL842.2),
 there are  disagreements between these values. For GL426 and 
GL842.2, the kinematics and other information seem to 
indicate that they are not  associated with the moving group. The case of
GL564.1 is discussed below.

Figures 2a and 2b allow us to identify possible members based on their 
kinematic  properties (V against U and W, respectively, from Hipparcos data).
Soderblom \& Mayor (1993) pointed out that the components of a
stellar kinematic group must have the same motion in the direction of the
Galactic rotation (the V component).
 Visual inspection indicates that the velocities are clustered around
 (-10,-8,-10). The dispersions 
of these velocities are rather small, in particular for the the 7
A spectral type stars (note that Castor is itself composed of two pairs of
spectroscopic binaries, but we have counted them as one star).
 However, there are several stars which have velocities
in disagreement with these average values.

In particular, the Galactic 
velocities (specifically, using the V component as a criterion)
 seem to indicate that GL426AB, GL466 and GL696 are probable 
non-members of the  moving group, and that GL803 and GL842.2
are possible non-members
(indicated with the labels ``N'' and ``N?'' in the second column of Table 4).
The table also indicates if the stars are probable or possible members
 (``Y'' and ``Y?'', respectively). 

GL564.1 ($\alpha$$^2$ Lib) deserves special attention.
It is physically associated to GL563.4
(Poveda et al. 1994). The WEB value of the radial velocities of both
stars leads to a a similar value of the V component of the Galactic velocity,
but Poveda et al. (1994) 
provide a radial velocity in large disagreement with the WEB
value. We have preferred to keep this last value.
 On the other hand, the U component of the velocity of GL563.4
 is quite different from  the average of the
group. The only relevant difference between both stars
is the radial velocity and this could be due to the orbital motion as well
as to  uncertainties in the measured values. Therefore, we do not rule out 
the possibility that it indeed  belongs to the moving group.

Figure 3c displays the velocities computed with the PPM data. 
The agreement in  the kinematics of these stars also suggest a 
common origin.  However, when computed
from the PPM values, Castor has a V component of the velocity quite 
different from the average of the moving group, with a difference much
larger than the quoted uncertainties. Therefore, it would be possible that
these system would not be a member of the moving group (and the association
should be called Vega moving group). In fact, if a very restricted definition
would be used, this figure indicates that there could be  2 different
groups; one including Vega, Fomalhaut and $\kappa$ Phe and another with 14 Lep
 and $\alpha^2$ Lib. Alderamin could be in either of them, due to its large
error bars. However, the data based on Hipparcos and the evidences
provided below seem to indicate that, indeed, all of them share a common
origin.

\subsection{Color-Magnitude Diagrams}

Further information on the membership can be obtained from Color-Magnitude
Diagrams. Figure 3a displays the absolute visual magnitude against
the (B--V) color, whereas Figure 3b has (V--I)$_c$ in the x-axis. 
The initial sample of late spectral type stars are shown as circles.
Filled symbols indicate final members and open symbols those stars rejected 
as physically associated (see Section 3.4).
Both  figures include the D'Antona and Mazzitelli (1994)
 isochrones for the ages
3, 10, 35 and 70 Myr (top to bottom, dashed lines). The solid line is the ZAMS.
Details concerning the convertion between 
 theoretical values (T$_{\rm eff}$ and luminosity) and the
 observational plane can be found in Stauffer et al. (1995) and
Barrado y Navascu\'es et al. (1997a).

We have selected those stars located below the  35 Myr isochrone
and above the ZAMS  as possible
members and indicated this in the third and fourth columns of Table 4.

\subsection{Stellar activity}

Figure 4 compares the equivalent width  of H$\alpha$  for our sample of stars 
(filled circles) with the Hyades (open circles). This graph can be used to 
support  rejection of several candidates or to accept them as members:
 GL696 presents a minimum value
in its H$\alpha$ absorption (it is a very inactive star), so it seems
to be older than the Hyades and a non-member of the moving group.
 GL896A, GL896B and GL803 are very active stars. Their activity is similar
to the most active Hyades stars, and indication that they could be younger.
However, Barrado y Navascu\'es et al. 1998) have shown that some dM Praesepe
stars have larger activity than equivalent Hyades stars (both clusters 
are coeval). Therefore, these stars could be as old as 600-800 Myr.
The spectroscopic binary YY Gem is a BY Dra system, whose rotational period is
synchronized with its orbital one, phenomenon which produces an enhanced
stellar activity, not related with the age of the system. Its location
 on the plot cannot provide any information about its membership to the 
moving group.  Similar conclusions can be reached if
 the coronal activities   are compared.

\subsection{A list of members}

We have combined the information provided by the kinematics and the CMD
to select a list of members. We have also taken into account the lithium 
abundance and the H$\alpha$ equivalent width
 when that information is available. We find that there is a good 
agreement between these criteria (if the kinematics are different from the
average value, then the location in the CMD usually desagree with membership).

Column six in Table 4 provides our final classification for the membership.
The hotter stars are classified as members based only on the kinematic.
However, they are relatively young, and the probability of having several
stars of A spectral type, leaving the Main Sequence, with the same 
spatial velocity, and not having been born at the same time,
 is quite low. We have a total of  16 probable members and 
another star which could be part of the group ($\alpha$$^1$ Lib).

\subsection{Are these stars physically associated?}

A significant fraction of the stars studied here fulfill the Soderblom \& Mayor
(1993) definition of a stellar kinematic group. The most restrictive 
requirement is that of having 
 identical V component of the Galactic velocity. In our case, 
there are 8 stars with V within 1 sigma of the average of  
the final members. This group includes Castor, Vega and Fomalhaut. Although
we do not rule out the possibility of having a spurious kinematic group, we
believe that the data presented here provide strong indications for a
common origin of, at least, the majority  of the sample.
However, the membership of individual stars cannot be known without doubt.
In fact, Whitmire et al. (1992) studied the kinematics of several A stars in 
the solar vicinity, including several stars in common with our study, 
namely, GL20, GL217.1, GL278, GL564.1, GL721, GL826 and GL881. Although they 
discussed the possibility of a common origin for several of these stars, 
they concluded that the small velocity dispersion is due to an observational
selection. Their model implies an interaction with 
an interstellar cloud, a 
phenomenon  which could enhance the circumstellar disks.
However, not all A stars in our sample display IR excesses, a subproduct 
of the disks. In any case, some caveats about particular membership 
should be kept in mind.

\section{Estimating the age of the moving group}

\subsection{Age and kinematics}

Palou$\check s$ \& Piskunov (1985) have shown that there is a relation between
the kinematic properties and the age of B and A stars. Based on their
Table 3, it is possible to constrain the age of the hot members of the 
Castor moving group. They have average values of the velocities 
of $<$U$>$=--10.7$\pm$3.5, $<$V$>$=--8.0$\pm$2.4 and 
$<$W$>$=--9.7$\pm$3.0 km/s, 
leading to ages of 100--400, 200-600 and 50-100 Myr, respectively.
These ages are indicative, since the velocity dispersion components
in the Palou$\check s$ \& Piskunov (1985) study
are quite large and, in the last case, the W component of the velocity
of the Castor group is slightly out the range of velocities examined by 
Palou$\check s$ \& Piskunov (1985).
A combined age would be 100-400 Myr, in good
agreement with the values derived for the late spectral type stars 
(see below).

\subsection{Age from the isochrone fitting}

Figures 3a and 3b can provide an estimate of the age of the association.
The final set of members lies below the 35 Myr isochrone and  above the 70 Myr.
Both the proximity to the ZAMS and the unknown
metallicity of the group do not allow an accurate determination of the  age.
However,  it seems  that  these stars are clearly above the ZAMS. Then, 
they would be Premain sequence (PMS) objects, still approaching at the MS. 
We believe that due to the errors in the conversion from the 
theoretical plane to observational plane, together with the errors associated
with the observational data (very small in most cases),
 the two color-magnitude diagrams  put only  a lower limit to
the age, of 35 Myr.

On the other hand, the hotter real members of the moving group
should be  close to  the end of their lives in the Main Sequence (in
particular, Castor
A and B, and $\alpha$$^2$~Lib). This particular evolutionary
status  allows for the possibility of  isochrone fit.
Again, uncertainties in the individual  actual distances --negligible, see
Table 3--,  magnitudes and colors are
too large  to provide accurate ages.
Backman \& Paresce (1993), by fitting isochrones, derived ages close
to 400 and 200 Myr for Vega and Fomalhaut, respectively, with 
uncertainties of 30\%. For our data set, we use
Meynet et al. (1993) set of evolutionary tracks. We find  that
Fomalhaut is too close to the MS to fit an isochrone. The location
of Vega, hotter than Fomalhaut,
 in the CMD is completely compatible with the turn-off position
of  the open cluster NGC~6494, and provides an age close to 300 Myr.  
 In any case, it seems that at least some
stars, including the Castor system, YY~Gem, Fomalhaut and Gl~879,
are physically associated and, for this reason, they have the
same age.

   The common origin and age of these stars would have important
implications. In particular,  YY Gem, an  eclipsing binary, is one
of the two M dwarfs with accurate measurements of  radii and
temperatures.
Chabrier \& Baraffe (1995), using new evolutionary models,  fit
isochrones to both components of the system (either in the radii-mass
or the T$_{\rm eff}$-mass planes), and  concluded that the system has to be
on the late Pre-main Sequence contraction phase. They estimated an age of 
$\sim$100 Myr. We note that  if it is really
 slightly a PMS system, then it should not be used as a calibrator
of theoretical models  (unless one can
accurately correct for its PMS nature or accurately know its age).

\subsection{The lithium--Teff plane}

The lithium abundance of late spectral type stars can be used as a 
membership criterion, as well as a statistical method to estimate the
age of an association of stars, since the lithium abundance depends
on both age and mass (e.g., Balachandran 1994).
 Figures 5a and 5b depict the abundance
against the effective temperature for those stars with published
values (crosses). In the first case (Fig. 5a),
 we have included data corresponding to Hyades stars
(filled circles) and Pleiades stars (open circles). 
 These two clusters have standard
ages of 600-800 and 70-80 Myr, respectively (see Stauffer et al. 1998 for a
new determination of the age of the Pleiades, which yields 125$\pm$5 Myr,
based on the position of the lithium boundary for very low mass stars and
brown dwarfs). The second figure  contains data from the  M34 cluster (open
circles) and the UMa moving group (filled circles), which are 200 and 300 Myr
 old,  respectively. Figure 5b also includes a 300 Myr lithium depletion
isochrone from Chaboyer (1993). 
The sources of the lithium data for all four associations can be found
in Barrado y Navascu\'es et al. (1997a).

The comparison between the lithium abundances of cluster stars of different
ages and GL879, a physical companion of Fomalhaut, was used by 
Barrado y Navascu\'es et al. (1997a) to establish that these stars are
younger than 300 Myr, probably in the range 100-300 Myr. Figure 5a and 5b 
indicates that the lithium abundance of $\alpha^1$ Lib is compatible
with this age (see our caveats about the membership based on the 
kinematic, section 3.5).
 GL848 seems too old to be part of the of the moving group,
 whereas GL896A is too young.

The case of YY Gem is more complex. 
It is well known that it is a very close SB2 eclipsing
binary. Its rapid rotation (v~sini=40 kms$^{-1}$)
causes significant for the important spectral line blending.
Moreover, since the system is composed of 2 similar stars
(M=0.62 and 0.57 M$_\odot$, Bopp 1974),  lines arising from both
components appear in the spectrum. For the particular
observation used to derive the lithium abundance (phase=0.21,
Barrado y Navascu\'es 1996), the LiI6708.8 \AA{ } doublets of each
component were well detached.
 The final abundance is
 Log N(Li)=0.02$\pm$0.20, in the customary scale where Log N(H)=12.
However, close binaries in clusters (Barrado y Navascu\'es \&
Stauffer 1996 in the case of Hyades, and Ryan \& Deliyannis 1995
 in the case of M67)
and chromospherically active binaries, such as YY Gem (Barrado y Navascu\'es
et al. 1997b) inhibit partially the lithium depletion, probably due
to mixing related to rotation. Therefore, this binary is not a good
candidate to be used to estimate the age of the Castor moving group,
although it is  very well established that there is a
 physical connection between these stars.

\subsection{The age of the association}

All these comparisons, which are compatible with 
 the well determined  age of GL879  and Fomalhaut, 
allows us to conclude that, indeed, the Castor moving group 
has an age of 200$\pm$100 Myr. Note that at this point, we have assumed
that the stellar kinematic group is real and that these stars are members
 of it. However, as  mentioned before,  the group, as found in
open clusters,  could contain spurious members, and even the possibility
of having a sample with the same kinematic properties and compatible CMD
 and  a different origin cannot be rejected.

If these stars do have a  common origin, as it seems,
 the same age derived for them is  a very relevant fact, since
the group has several A stars with different infrared excesses
(or none at all), indicating that, at least at this age, the
evolution of the protoplanetary disks does not depend
on their age.

\section{Summary}

We have used data provided by Hipparcos to select several stars as possible
 members of the 
Castor moving group. Additional information,
such as the position in Color-Magnitude Diagrams, isochrones, 
lithium abundances, stellar activity, was used to add new constraints
to the membership and to estimate the age of the association,
which is 200$\pm$100 Myr.

The moving group contains seven A spectral type stars with different
IR excesses. This fact could indicate
that, at this stage, the evolution of these excesses,
when present, does not depend on age, and that, 
therefore, the time scale of the
formation of planetary does not seem to be universal.
The excesses would be produced by collisions and sublimation of
large bodies and  not from protoplanetary disks.


\acknowledgements{
This research has made use of the Simbad database, operated at
CDS, Strasbourg,  France.
DBN acknowledge the support by 
the ``Real Colegio Complutense at Harvard University'' and the
MEC/Fulbright Commission,  and the very useful comments and suggestions
by the referee, Dr. Soderblom.}

\begin{center}
{\sc Figure Captions}
\end{center}

{Figure 1.- Color-Magnitude Diagram for the initial sample of proposed members
of the Castor moving group. The Schmidt-Kaler ZAMS is included as a solid 
line.}

{Figure 2.- kinematics of the whole sample. The seven proposed members of
A spectral type are indicated with filled circles and labels. The c panel 
shows the data obtained using the PPM data}

{Figure 3.- Color-Magnitude Diagrams for the late spectral type candidates.
The isochrones (3, 10, 35, 70 and ZAMS) 
are those from D'Antona \& Mazzitelli (1994). Final members
are indicated as filled circles, whereas rejected members appear as open 
circles.
{\bf a} Mv against (B--V).
{\bf b} Mv against (V--I)$_c$.
}

{Figure 4.- H$\alpha$ equivalent width against  (V-I) color. Proposed members
of the moving group are shown as filled circles. Hyades members appear as 
open circles.}

{Figure 5.- Lithium abundances against effective temperatures. Crosses 
represent the proposed members of the moving group. 
{\bf a} Pleiades and Hyades data (open and filled circles). 
{\bf b} M34 and UMa Group  data (open and filled circles). 
}

\end{document}